\documentclass[aps,prl,twocolumn,showpacs,groupedaddress]{revtex4}
\usepackage{graphicx}	% Include figure files
\usepackage{dcolumn}	   % Align table columns on decimal point
\usepackage{bm}	      % bold math
\usepackage{ulem}
\usepackage{lscape}

\begin{document}

\include{epsf}

%\preprint{}

\title{Feshbach resonances in ultracold $^{6,7}$Li + $^{23}$Na atomic mixtures}



\author{Marko Gacesa, Philippe Pellegrini, and Robin C\^ot\'e}

%\thanks{}
\affiliation{Department of Physics, U-3046, University of Connecticut, 
Storrs, CT, 06269-3046}


\date{\today}

\begin{abstract}
We report a theoretical study of Feshbach resonances in $^{6}$Li + $^{23}$Na 
and $^{7}$Li + $^{23}$Na mixtures at ultracold temperatures using
new accurate interaction potentials in a full quantum coupled-channel 
calculation. Feshbach resonances for $l=0$ in the initial collisional open channel 
$^6$Li$(f=1/2, m_f=1/2) + ^{23}$Na$(f=1, m_f=1)$ are found to agree 
with previous measurements, leading to precise values
of the singlet and triplet scattering lengths for the $^{6,7}$Li$+^{23}$Na 
pairs. We also predict additional Feshbach resonances within experimentally
attainable magnetic fields for other collision channels.
\end{abstract}

%  \pacs{32.80.Pj}
% \keywords{}

\maketitle

A Feshbach resonance occurs when the kinetic energy of two colliding 
atoms matches the energy of a bound level associated with a closed 
channel \cite{Feshbach:1992}. Since the first experimental observation 
in a Bose-Einstein condensate (BEC) of sodium \cite{Nature.392.151}, 
Feshbach resonances have been widely used as a tool for controlling 
interactions in ultracold atomic gases. One can vary the sign and 
strength of the interactions by tuning the 
magnetic field near such a resonance, allowing the study of ultracold 
atomic gases from the strong interacting to the non-interacting 
regime \cite{PhysRevA.47.4114}. For instance, BECs of
$^{133}$Cs \cite{TinoWeber01102003} or
$^{85}$Rb \cite{PhysRevLett.85.1795} have been realized using this
technique to reverse the sign of the scattering length from negative
to positive. Furthermore, Feshbach resonances can be used as a
spectroscopic tool \cite{PhysRevLett.85.2721},
and provide a way of making ultracold molecules \cite{Nature.424.47}.
They allow for the coherent coupling between atomic and molecular 
condensates \cite{PhysRevLett.83.2691}, as well as the creation 
of molecular BECs \cite{Nature.426.537}.
More details can be found in Ref. \cite{kohler:1311}

Recently, Feshbach resonances have been observed in Bose-Fermi
mixtures of $^6$Li and $^{23}$Na \cite{stan:143001} as well as in 
 $^{87}$Rb and $^{40}$K \cite{inouye:183201}, opening 
the door to new phenomena such as boson-mediated Cooper 
pairing \cite{PhysRevA.61.053601} or the formation of heteronuclear 
molecules \cite{stirap-KRb}. Such polar molecules may have applications in quantum 
computation \cite{PhysRevLett.88.067901}, as well as in the search for the 
electronic dipole moment \cite{PhysRevLett.89.133001}, 
or study of degenerate gases with dipolar interactions \cite{PhysRevA.66.013606}.
The most common theoretical treatment of cold collisions is the coupled-channel 
calculation approach \cite{PhysRevB.38.4688}, although several simplified 
methods have been developed \cite{nygaard:042705}. These techniques were 
applied to K-Rb mixtures by Simoni \textit{et al.} \cite{PhysRevLett.90.163202} 
and to mixed-isotope mixtures of rubidium by Burke \textit{et al.} \cite{PhysRevLett.80.2097}.

In this article we present an extensive theoretical study of the 
scattering properties of $^{6,7}$Li and $^{23}$Na in the presence 
of a magnetic field and at the temperature range typical of degenerate 
gases (i.e. between $300$ nK and $1 \mu$ K). Since Li has both 
fermionic and bosonic isotopes, Li-Na mixtures are of particular 
interest. We focus our efforts on finding the positions and widths 
of Feshbach resonances for various collision channels. This 
information will be useful for future experiments on cold Li-Na 
mixtures, {\it e.g.} to form a BEC of polar molecules.

For two alkali atoms of relative momentum $\vec{p}$ and reduced mass 
$\mu$ colliding in a magnetic field, the effective two-body Hamiltonian 
can be written as \cite{PhysRevA.51.4852}
\begin{equation}
  \label{Hamiltonian}
  H=\frac{p^2}{2\mu}+\sum_{j=1}^{2}H^{\rm int}_j +V^c+V^d \;,
\end{equation}
where $H^{\rm int}_j$ is the internal energy 
of atom $j$, $V^c$ the Coulomb interaction, and $V^d$ the magnetic 
dipole-dipole interaction (often neglected \cite{kohler:1311}).
$H^{\rm int}_j$ consists of the hyperfine and Zeeman
contributions, respectively
\begin{equation}
  H^{\rm int}_j=\frac{a^{(j)}_{\rm hf}}{\hbar^2}\vec{s}_j\cdot\vec{i}_j + 
  (\gamma_e\vec{s}_j - \gamma_n\vec{i}_j)\cdot \vec{B} \; .
\end{equation}
Here $\vec{s}_j$ and $\vec{i}_j$ are the electronic and nuclear spin of atom $j$, 
$a^{(j)}_{\rm hf}$ its hyperfine constant (152.1368407 MHz for $^6$Li, 
401.752 MHz for $^7$Li and 885.813 MHz for $^{23}$Na), and
$\vec{B}$ is the external magnetic field assumed in the $z$-direction.

The Coulomb interaction, which depends only on the internuclear 
separation $R$, can be decomposed into singlet and triplet 
contributions
\begin{equation}
  V^c=V_0(R)P^0+V_1(R)P^1 \;,
\end{equation}
where $V_0$ ($V_1$) and $P^0$ ($P^1$) stands for the singlet (triplet) molecular potential 
and its associated projection operator respectively \cite{Feshbach:1992}. 
Molecular potentials were constructed by smoothly connecting the 
inner part to the long-range dispersion expansion 
\begin{equation}
   V_{\rm LR} = -\frac{C_6}{R^6}-\frac{C_8}{R^8}-\frac{C_{10}}{R^{10}} \pm 
                  C_{\rm ex} e^{-b R}
\end{equation}
where $\pm$ are for the triplet and singlet potential respectively. We used the 
singlet potential of Fellows \cite{fellows:5855}, constructed from 
accurate spectroscopic data via the inverse perturbation approach (IPA), and 
an {\it ab initio} triplet potential computed using the \textit{CIPSI} 
package \cite{aymar:204302} and adjusted to match the atomic
dissociation energy of Li(2s) + Na(3s) at infinity. For the long range
form of both potentials, we adopted the dispersion coefficients ($C_{\rm n}$) and 
exchange energy ($C_{\rm ex}$ and $b$) of Fellows (case a) \cite{fellows:5855}.

At ultracold temperatures, $s$-wave collisions describe the scattering process. 
The collision entrance channel is determined by the initial Zeeman states of the 
atoms. A judicious choice of basis is the field dressed molecular hyperfine states
\cite{PhysRevB.38.4688}: $|{\alpha \beta} \rangle$, with 
$\alpha=1,\dots ,6$ for $^6$Li, $\alpha=1,\dots ,8$ for $^7$Li, and 
$\beta=1,\dots ,8$ for $^{23}$Na, all in the increasing order of energy.
We follow this standard labeling throughout the article.
In the limit $B=0$, it reduces to the product of atomic hyperfine states
\begin{equation}
  \label{Vdd2}
  |f_1,m_{1};f_2,m_{2}\rangle\equiv|f_1,m_{1}\rangle_{\rm Li}\otimes
  |f_2,m_{2}\rangle_{\rm Na}
\end{equation}
where $\vec{f}_j=\vec{s}_j+\vec{i}_j$ is the total spin of atom $j$, 
and $m_j$ its projection onto the molecular axis. This basis is suitable 
for description of the collision in the limit of two 
separated atoms. At smaller separation, the molecular basis 
$|SIFM_F\rangle$ becomes more appropriate, with 
$\vec{S}=\vec{s}_1+\vec{s}_2$, $\vec{I}=\vec{i}_1+\vec{i}_2$, 
$\vec{F}=\vec{f}_1+\vec{f}_2$, and $M_F$ its projection.

Once the bases are defined, one can express Eq.(\ref{Hamiltonian}) 
in matrix form, leading to the following matrix equation
\begin{equation}
  \label{cc}
  \frac{d^2}{dR^2}\textbf{F}(R)=\frac{2\mu}{\hbar^2}\textbf{M}(R)\textbf{F}(R) ,
\end{equation}
where the coupling matrix $\textbf{M}(R)$ is defined as
\begin{equation}
  \textbf{M}_{(\alpha,\beta)}^{(\alpha',\beta')}(R)\frac{\delta(R-R')}{RR'}
  = \langle R(\alpha,\beta)|V^c|R'(\alpha',\beta') \rangle ,
\end{equation}
and $\textbf{F}(R)$ is a matrix with columns corresponding to a 
complete set of linearly independent solutions. 

To obtain results with the level of accuracy given in the
experimental measurements of Stan {\it et al.} \cite{stan:143001},
we needed to include the nuclear Zeeman term $\gamma_n\vec{i}_j\cdot \vec{B}$ and the magnetic dipole-dipole interaction term $V^d$ \cite{note-Vd}.
The latter one can be written as
\begin{eqnarray}
  \label{Vdd}
  V^d & = & -\alpha_{\rm fs}^2 \frac{3(\hat{\textbf{R}} \cdot \hat{\textbf{s}}_1)
             (\hat{\textbf{R}} \cdot \hat{\textbf{s}}_2)-
             \hat{\textbf{s}}_1 \cdot \hat{\textbf{s}}_2 }{R^3} \nonumber \\
      & = & -\frac{\alpha_{\rm fs}^2 \sqrt{6}}{R^3} 
             \sum_{q=-2}^{2}{(-1)^q C_q^2(\vec{s_1} \otimes \vec{s_2})_{-q}^{(2)}},
\end{eqnarray}
where $\alpha_{\rm fs}$ is the fine structure constant. 
In the second relation, the spin and orbital part are separated,
with $C_q^2(\theta,\phi)$ representing a reduced spherical harmonic 
operator and $(\vec{s_1} \otimes \vec{s_2})_{-q}^{(2)}$ the second rank 
tensor operator that couples the spins \cite{Zare}. We took into account 
the contribution from $V^d$ by expanding the bases to include the 
total orbital angular momentum $l$ and its projection $m_l$: 
$|l m_l\rangle$. We calculated the matrix elements of Eq.(\ref{Vdd}) 
in the fully uncoupled basis $|l m_l;\lambda\rangle$, where
$\lambda$ stands for set of quantum numbers 
$\{s_1, m_{s_1}, i_1, m_{i_1}, s_2, m_{s_2}, i_2, m_{i_2}\}$. With this basis,
$\langle \lambda|V^d|\lambda '\rangle$ takes the simple form \cite{Krems:2296}

\begin{eqnarray}
\label{Vdd3}
 \delta_{\lambda \lambda '} &-&  (-1)^{-m_l}
   \frac{\sqrt{6} \alpha_{\rm fs}^2}{R^3}\sqrt{(2l+1)(2l'+1)} \nonumber \\ 
   & \times &\left( \begin{array}{ccc}
       l & 2 & l' \\
      0 & 0 & 0
      \end{array} \right)
   \sum_{q=-2}^{2} { (-1)^q \left(
   \begin{array}{ccc}
    l & 2 & l' \\
    -m_l & -q & -m_l'
  \end{array} \right) } \nonumber \\
 & \times &  \langle s_1 m_{s_1}| \langle s_2 m_{s_2} | [\vec{s_1} 
    \otimes \vec{s_2}]_q^{(2)} 
    | s_1 m_{s_1}' \rangle |s_2 m_{s_2}' \rangle .
\end{eqnarray}
Non-zero matrix elements for $\Delta l = 0, \pm 2$ and 
$l = l' \neq 0$ couple $s$ and $d$ waves.  We
included small corrections to the energies from higher 
partial waves up to $l=4$.

We have also checked that there are no higher partial wave 
resonances ({\it e.g.}, $p$-wave, $l=l'=1$) for the channel 
$|11\rangle$.
Since the system remains invariant under rotations with respect 
to the axis of magnetic field, the projection
of the total magnetic quantum number $M_F=m_{1} + m_{2}$ is conserved, 
and it determines the number of coupled channels.
The Feshbach resonances listed in this work are all $s$-wave
resonances ($l=0$).

We solve Eq.(\ref{cc}) for different
values of magnetic field using the multichannel 
log-derivative method \cite{Johnson1973} to obtain 
the $\textbf{S}$-matrix, from which we extract the radial 
wave function phase shift $\eta(k)$ of the initial entrance
 channel, as well as the scattering length
$a$ using \cite{Mott:1987}
\begin{equation}
\lim_{k\rightarrow 0}k\cot \eta(k)=-\frac{1}{a} .
\end{equation}
Here, $k=\sqrt{2\mu E}/\hbar$ is the wave number associated 
with the pair of colliding atoms of relative energy $E$.

\begin{figure}[top]
  \centerline{\epsfxsize3.25in\epsfclipon\epsfbox{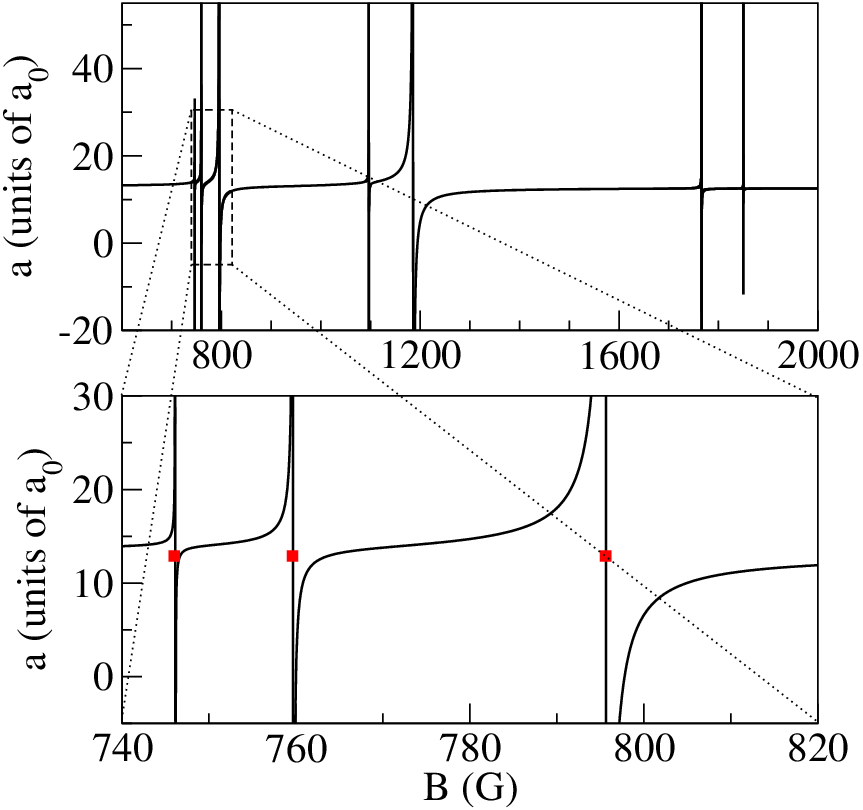}}
  \caption{\textit{Top:} Scattering length for $^6$Li+$^{23}$Na in 
  the entrance channel $|{11} \rangle$. Positions and widths of the resonances 
  are given in Table~\ref{table1}. \textit{Bottom:} Zoom on the first three
  resonances. Squares are the experimental positions reported in \cite{stan:143001}.}
\label{fig1}
\end{figure}

We first calculated the positions of Feshbach resonances 
for the least energetic hyperfine state $|\alpha \beta \rangle = 
|11 \rangle$ of $^6$Li+$^{23}$Na. There is no inelastic spin 
relaxation for this channel which makes it attractive for 
trapping since the only possible decay channel is through 
3-body interactions. Stan \textit{et al.} \cite{stan:143001} have measured
the positions of three resonances for this state and we used 
their results to calibrate the singlet and triplet
potentials. For each curve, we varied the inner wall by shifting
the positions of the data points for separations less than the 
equilibrium separation $R_e$ according to 
$R_{\rm shifted} = R+ s(R-R_e)/(R_c-R_e)$, where $s$ 
corresponds to the shift of the zero-energy classical turning
point $R_c$. The best agreement with \cite{stan:143001} 
was obtained with $s=0.06170$ a.u. (singlet)
and $s=-0.32878$ a.u. (triplet).
It may also be worth noting that we obtained a very good agreement 
with experimental results only after we included the
coupling with nuclear spin and magnetic dipole-dipole
interaction. Neglecting these second order terms
resulted in agreement up to $\pm5$ Gauss for the
measured resonances. Our results for the entrance channel 
$|11\rangle$ and the magnetic field up to 2000 Gauss are shown in 
Fig.~\ref{fig1} and summarized in Table~\ref{table1}.
Note that four additional Feshbach resonances were found 
at higher values of magnetic field, at 1097, 1186, 1766 and 1850 Gauss.

\begin{table}
\caption{\label{table1} $^{6}$Li+$^{23}$Na - calculated Feshbach resonances ($l=0$)
  for the entrance channel $|11\rangle$ and $B$ up to 2000 Gauss.
  We compare with the measured position \cite{stan:143001},
  and list the width $\Delta$ and the background scattering length $a_{\rm bg}$. 
  Both $a_{\rm bg}$ and $\Delta$ were obtained by fitting to 
  $a_0 = a_{\rm bg} \left(1 - \frac{\Delta}{B-B_0} \right)$. Partial waves $l=0 \dots 4$ 
were included in the 
  calculation.}
\begin{ruledtabular}
\begin{tabular}{cccc }
 $B_{0}^{\rm th}$ (G) & $B_{0}^{\rm exp}$ (G) \cite{stan:143001}
                  & $\Delta$ (G) & $a_{\rm bg}$ ($a_0$) \\
\hline
  746.13 & $746.0 \pm 0.4$ & 0.044  & 14.003 \\
  759.69 & $759.6 \pm 0.2$ & 0.310  & 13.864 \\
  795.61 & $795.6 \pm 0.2$ & 2.177  & 13.002 \\
 1096.68 &                 & 0.153  & 13.902 \\
 1185.70 &                 & 8.726  & 12.673 \\
 1766.13 &                 & 0.156  & 12.500 \\
 1850.13 &                 & 0.019  & 12.499 \\
\end{tabular}
\end{ruledtabular}
\end{table}

With this set of accurately adjusted potentials, 
we have also determined the singlet ($a_S$) and
triplet ($a_T$) scattering lengths and the energy of 
the last vibrational level ($E_{\rm last}^{S,T}$)
for both $^6$Li+$^{23}$Na and $^7$Li+$^{23}$Na mixtures 
(see Table \ref{table2}).
Stan \textit{et al.} \cite{stan:143001} report a good 
thermalization rate between for the $^6$Li+$^{23}$Na 
mixture and exploit the effect for efficient sympathetic
cooling. Our results are in agreement with the estimate 
based on those rates. From the thermalization rate in their experiment,
the MIT group estimated the ratio of pure triplet elastic collision
between Na-Na ($\sigma_{AA}$) and Li-Na ($\sigma_{AB}$) to be 
$\sigma_{AA}/\sigma_{AB}\sim 100$ \cite{Ketterle:private_comm},
where $\sigma_{AA}=8\pi a^2_{\rm Na-Na}$ and 
$\sigma_{AB}= 4\pi a^2_{\rm Li-Na}$, 
with $a_{\rm Na-Na}=85$ $a_0$ for the triplet scattering length of Na, 
one then estimates $|a_{\rm Li-Na}|=12$ $a_0$ for the triplet case.
This estimate is also in agreement with the value of
$|a_T| \approx 15$ $a_0$ obtained assuming a thermalization time 
$\tau_{\rm tot}\sim 4 \tau \sim 15$ sec., where the relationship 
between $\tau$ and $a$ is given in \cite{PhysRevLett.80.3419}.
The uncertainty in the scattering 
length was determined by adjusting the inner wall of the potentials 
to match the uncertainty of the resonances (Table \ref{table1}).
Alternatively, we shifted the triplet potential below the last bound state 
by $\pm 140$ MHz \cite{stan:143001}, and then 
adjusted the inner wall of both singlet and triplet curves to match
the experimental position of the resonances.
The last triplet bound level was then found with smaller
uncertainty to be $E_{\rm last}^T = -5720 \pm 16$ MHz, 
as compared to the estimated value of 
$E_{\rm last}^{T} = -5550 \pm 140$ MHz of 
Stan \textit{et al.} \cite{stan:143001}. 
The same procedure was repeated after varying $C_6$ coefficient
by $\pm 5\%$, and the results were within the uncertainties given in Table \ref{table2}.
Note that Feshbach resonances are very
sensitive to the energy of the last vibrational level, 
which thus can be determined very precisely,
while the total number of levels can only be estimated
since it depends on the whole potential, which is usually
not known accurately enough. For our adjusted triplet
state of $^6$Li$^{23}$Na, we found 11 vibrational levels
(see Table~\ref{table2}).
% (Table~\ref{table2} lists the energy and number of bound
% levels for all isotopomers, both singlet and triplet).

\begin{table}
\centering
\caption{\label{table2} $^{6,7}$Li+$^{23}$Na singlet ($S$) and triplet ($T$) scattering 
         lengths in units of $a_0$, the last vibrational level, and the corresponding 
         binding energy. See the text for the description of the uncertainty.}
\begin{ruledtabular}
\begin{tabular}{cccc}
  & \multicolumn{2}{c}{\vspace{0.1cm} Present work}  & Ref. \cite{PhysRevLett.80.3419} \\
  & $^7$Li+$^{23}$Na & $^6$Li+$^{23}$Na & $^6$Li+$^{23}$Na \\
\hline
   $a_S$ & $39.7 \pm 0.5 ~(1.6)$ & \textbf{$15.9 \pm 0.3 ~(1.5)$} & $39.2$  \\
   $a_T$ & $36.1 \pm 0.3 ~(4.6)$ & \textbf{$12.9 \pm 0.6 ~(4.5)$} & $31.1$  \\
   $v_{\rm last}^S$  & 47 & 45 & \\
   $v_{\rm last}^T$  & 12 & 11 & \\
   $E_{\rm last}^{S}$ (MHz) & $-1505 \pm 3 $ & $-1.6 \pm 0.2$ & \\
   $E_{\rm last}^{T}$ (MHz) & $-7112 \pm 12$ & $-5720 \pm 16$ & \\
\end{tabular}
\end{ruledtabular}
\end{table}

\begin{figure}
\centerline{\epsfxsize3.25in\epsfclipon\epsfbox{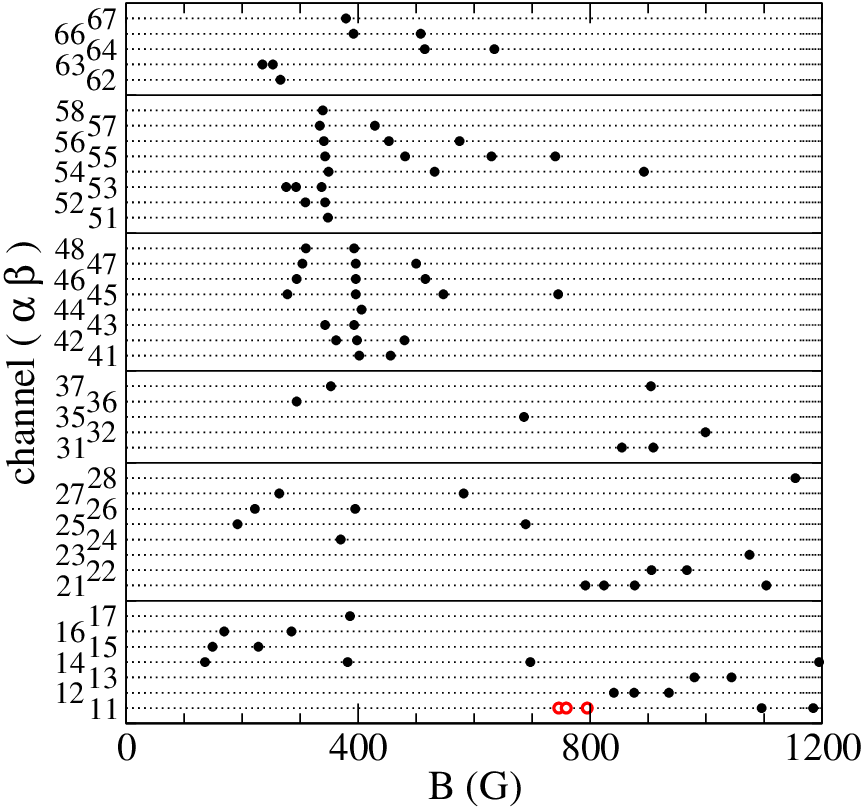}}
\caption{\label{fig2}Feshbach resonances in $^6$Li$+^{23}$Na for different collisional 
              entrance channels. Only channels for which the resonances were 
              found are shown. Open circles for channel $|11\rangle$ were observed experimentally.}
\end{figure}

\begin{figure}
\centerline{\epsfxsize3.25in\epsfclipon\epsfbox{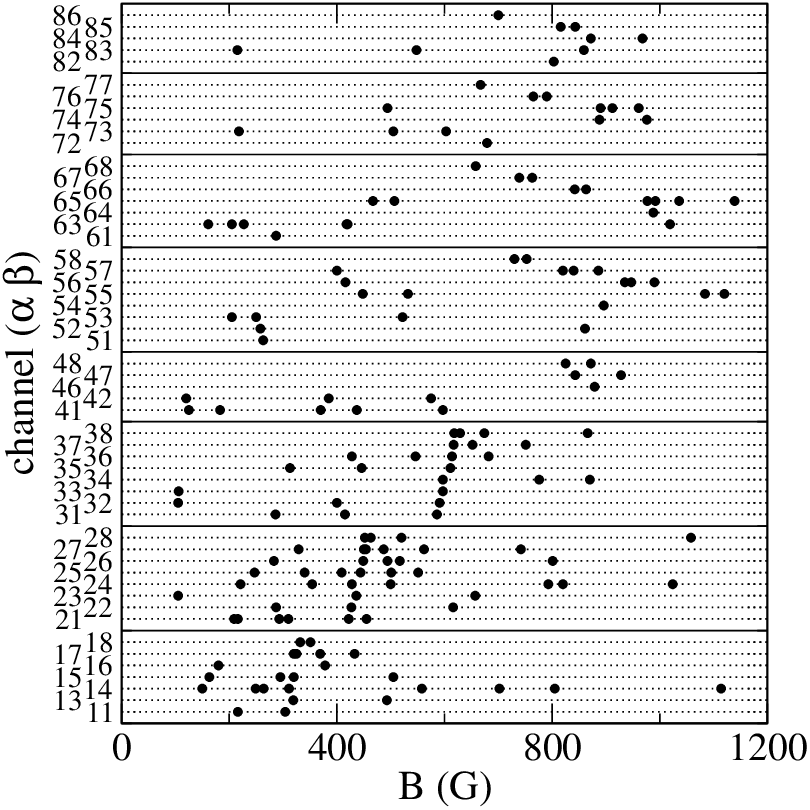}}
\caption{\label{fig3} Same as Fig.~\ref{fig2} for $^7$Li$+^{23}$Na.}
\end{figure}

Using our adjusted singlet and triplet potential curves, we performed
a similar calculation for all entrance channels for 
$^{6}$Li$+^{23}$Na and $^{7}$Li$+^{23}$Na.
Predicted positions of Feshbach resonances are shown schematically in Figures~\ref{fig2}
and \ref{fig3}. 
It appears that the $^7$Li+$^{23}$Na mixture has several experimentally 
attainable Feshbach resonances for the high-field seeking hyperfine states. 
Altogether, the richness of the scattering properties makes 
$^{6,7}$Li+$^{23}$Na mixtures interesting for further exploration.

In conclusion, we present an extensive study of scattering properties in 
ultracold $^6$Li+$^{23}$Na and $^7$Li+$^{23}$Na mixtures. A full 
quantum coupled-channel calculation was performed in the field dressed 
approach, to determine the positions and widths of several Feshbach resonances 
associated with different entrance channels. The accuracy of our singlet and triplet 
potentials improved by previous experimental measurements allowed 
us to very accurately determine the singlet and triplet scattering length 
of the system, as well as to give an estimated number of vibrational levels
in those potentials together with the energy of the last level.

Z. Pavlovic is gratefully acknowledged for providing the original log-derivative 
program. The authors are grateful to O. Dulieu and M. Aymar for providing us the 
LiNa \textit{ab initio} potentials, and W. Ketterle and V. Kharchenko for useful
discussions. This research was supported by the National Science Foundation.


\begin{thebibliography}{99}
\bibitem{note-Vd}
  The correction to the eigenenergies was at most
  1.5\% at small distance, and decreased rapidly as $R^{-3}$ with $R$.
\end{thebibliography}
\end{document}